\documentclass[prd]{revtex4}%
\usepackage{amsfonts}
\usepackage{amsmath}
\usepackage{amssymb}
\usepackage{graphicx}%

\begin{document}
\preprint{HEP/123-qed}
\title{Exact solutions of a Flat Full Causal Bulk viscous FRW cosmological
model\\
through factorization}
\author{O. Cornejo-P\'erez}
\affiliation{Facultad de Ingenier\'{\i}a, Universidad Aut\'onoma de
Quer\'etaro, Centro Universitario Cerro de las Campanas, 76010
Santiago de Quer\'etaro, Mexico}
\author{J. A. Belinch\'{o}n}
\affiliation{Departamento de F\'{\i}sica At\'{o}mica, Molecular y
Nuclear. Universidad Complutense de Madrid, E-28040 Madrid,
Espa\~{n}a}

\pacs{PACS number}

\begin{abstract}
We study the classical flat full causal bulk viscous FRW cosmological model
through the factorization method. The method shows that there exists a
relationship between the viscosity parameter $s$ and the parameter $\gamma$
entering the equations of state of the model. Also, the factorization method
allows to find some new exact parametric solutions for different values of the
viscous parameter $s$. Special attention is given to the well known case
$s=1/2$, for which the cosmological model admits scaling symmetries.
Furthermore, some exact parametric solutions for $s=1/2$ are obtained through
the Lie group method.


\end{abstract}

\date[Date text]{: \today}

\startpage{1}
\endpage{ }
\maketitle

\section{Introduction.}

Factorization of linear second order differential equations is a well
established method to find exact solutions through algebraic procedures. It
was widely used in quantum mechanics and developed since Schrodinger's works
on the factorization of the Sturm-Liouville equation. At the present time,
very good informative reviews on the factorization method can be found in open
literature (see for instance \cite{mielnik,rosu2}). However, in recent times
the factorization method has been applied to find exact solutions of nonlinear
ordinary differential equations (ODE) \cite{berkovich,
cornejo1,wang1,cornejo2,estevez}. In \cite{cornejo1}, based on previous
Berkovich's works \cite{berkovich}, it has been provided a systematic way to
apply the factorization method to nonlinear second order ODE. In \cite{wang1},
Wang and Li extended the application to more complex nonlinear second and
third order ODE. The factorization of some ODE may be restricted due to
constraints which appear in a natural way within the factorization procedure.
However, here it is shown that by performing transformation of coordinates,
one can be able to get exact parametric solutions of an ODE which does not
allow its factorization or presents cumbersome constraints.

The purpose of the present work is to apply the factorization method to study
the full causal bulk viscous cosmological model with flat FRW symmetries.
Since the Misner \cite{Mi66} suggestion stressing the fact that the observed
large scale isotropy of the Universe may be due to the action of the neutrino
viscosity when the Universe was about one second old, there have been numerous
works pointing out the importance of the physical processes involving viscous
effects in the evolution of the Universe (see for instance \cite{ChJa96}). Due
to such assumption, dissipative processes are supposed to play a fundamental
role in the evolution of the early Universe.

The theory of relativistic dissipative fluids, created by Eckart \cite{Ec40}
and Landau and Lifshitz \cite{LaLi87} has many drawbacks, and it is known that
it is incorrect in several respects mainly those concerning causality and
stability. Israel \cite{Is76} formulates a new theory in order to solve these
drawbacks. This theory was latter developed by Israel and Stewart
\cite{IsSt76} into what is called transient or extended irreversible
thermodynamics. The best currently available theory for analyzing dissipative
processes in the Universe is the full causal thermodynamics developed by
Israel and Stewart \cite{IsSt76}, Hiscock and Lindblom \cite{HiLi89} and
Hiscock and Salmonson \cite{HiSa91}. The full causal bulk viscous
thermodynamics has been extensively used to study the evolution of the early
Universe and some astrophysical process \cite{HiLi87,Ma95}.

The paper is organized as follows. In Section II, we start by reviewing the
main components of a flat bulk viscous FRW cosmological model, and introduce
the factorization technique as applied to the cosmological model. Field
equations (FE) of the classical bulk viscous FRW cosmological model
\cite{Ma95} reduce to a single nonlinear second order ODE, the fundamental
dynamical equation for the Hubble rate. By performing a transformation of both
the dependent and independent variables and using the factorization method,
this equation is transformed into a nonlinear first order ODE. The order
reduction of the equation for the Hubble rate allows to find a variety of new
exact parametric solutions of the FE for the viscous FRW cosmological model.
Furthermore, the factorization technique provides relationships for parameters
entering the factorized equation. Then, a noteworthy result is that the
viscosity parameter $s$ is not longer assumed to be independent of the values
of parameter $\gamma$. Such parameter relationships have not been previously
reported. In Section III, several particular models for $s\neq1/2$ are
studied. We obtain new exact parametric solutions through factorization and
compare with the ones obtained by several authors
\cite{C1,Ch97,H1,H2,H3,H4,H5,H6} who use different approaches. Section IV is
devoted to the special case $s=1/2$, for which the model admits scaling
symmetries. The scaling solution, previously studied by many authors is
obtained. In order to obtain more new solutions and compare the solutions
obtained through factorization for $s=1/2$, we consider the Lie group method
for this special case in Section V. Some conclusions end up the paper in
Section VI.

\section{The model.}

We consider a flat FRW Universe with line element
\begin{equation}
ds^{2}=-dt^{2}+f^{2}(t)\left(  dx^{2}+dy^{2}+dz^{2}\right)  , \label{1}%
\end{equation}
where the energy-momentum tensor of a bulk viscous cosmological fluid is given
by \cite{Ma95}:
\begin{equation}
T_{i}^{k}=\left(  \rho+p+\Pi\right)  u_{i}u^{k}+\left(  p+\Pi\right)
\delta_{i}^{k}, \label{2}%
\end{equation}
where $\rho$ is the energy density, $p$ the thermodynamic pressure, $\Pi$ the
bulk viscous pressure and $u_{i}$ the four-velocity satisfying the condition
$u_{i}u^{i}=-1$. We use the units $8\pi G=c=1$. The gravitational field
equations together with the continuity equation, $T_{i;k}^{k}=0,$ are given as
follows%
\begin{align}
2\dot{H}+3H^{2}  &  =-p-\Pi,\label{fe1}\\
3H^{2}  &  =\rho,\\
\Pi+\tau\dot{\Pi}  &  =-3\xi H-\frac{1}{2}\tau\Pi\left(  3H+\frac{\dot{\tau}%
}{\tau}-\frac{\dot{\xi}}{\xi}-\frac{\dot{T}}{T}\right)  ,\\
\dot{\rho}  &  =-3\left(  \gamma\rho+\Pi\right)  H, \label{fe4}%
\end{align}
where $H=\dot{f}/f.$ In order to close the system of equations we are assuming
the following equations of state \cite{Ma95}
\begin{equation}
p=\left(  \gamma-1\right)  \rho,\quad\xi=\alpha\rho^{s},\quad T=\beta\rho
^{r},\quad\tau=\xi\rho^{-1}=\alpha\rho^{s-1}, \label{steq1}%
\end{equation}
where $T$ is the temperature, $\xi$ the bulk viscosity coefficient and $\tau$
the relaxation time. The parameters satisfy $\gamma\in\left[  1,2\right]  ,$
$s\geq0$, and $r=\left(  1-\frac{1}{\gamma}\right)  $. The growth of entropy
has the following behavior%
\begin{equation}
\Sigma\left(  t\right)  \thickapprox-3k_{B}^{-1}\int_{t_{0}}^{t}\Pi
Hf^{3}T^{-1}dt.
\end{equation}

The Israel-Stewart-Hiscock theory is derived under the assumption that the
thermodynamical state of the fluid is close to equilibrium, i.e., the
non-equilibrium bulk viscous pressure should be small when compared to the
local equilibrium pressure $|\Pi|<<p=(\gamma-1)\rho$. Then, we may define the
$l(t)$ parameter as: $l=|\Pi|/p.$ If this condition is violated then one is
effectively assuming that the linear theory also holds in the nonlinear regime
far from equilibrium. For a fluid description of the matter, the condition
ought to be satisfied.

To see if a cosmological model inflates or not it is convenient to introduce
the deceleration parameter $q=dH^{-1}/dt-1$. The positive sign of the
deceleration parameter corresponds to standard decelerating models, whereas
the negative sign indicates inflation.

The fundamental dynamical equation for the Hubble rate is given by
\cite{Ma95}
\begin{equation}
\ddot{H}-A\frac{\dot{H^{2}}}{H}+\left(  3H+CH^{2-2s}\right)  \dot{H}%
+DH^{3}+EH^{4-2s}=0, \label{eq1}%
\end{equation}
where
\begin{equation}
A=\left(  1+r\right)  =2-\frac{1}{\gamma},\quad B=3,\quad C=3^{1-s},\quad
D=\frac{9}{4}\left(  \gamma-2\right)  ,\quad E=\frac{1}{2}3^{2-s}\gamma.
\end{equation}

Let us perform the following transformation of the dependent and independent
variables
\begin{equation}
H=y^{1/2},\qquad d\eta=y^{1/2}dt, \label{cv1}%
\end{equation}
then Eq. (\ref{eq1}) turns into
\begin{equation}
\frac{d^{2}y}{d\eta^{2}}-\frac{A}{2y}\left(  \frac{dy}{d\eta}\right)
^{2}+\left(  3+Cy^{\frac{1}{2}-s}\right)  \frac{dy}{d\eta}+2y(D+Ey^{\frac
{1}{2}-s})=0. \label{eq2}%
\end{equation}

Let us consider now the following factorization scheme \cite{cornejo1,wang1}.
The nonlinear second order equation
\begin{equation}
y^{\prime\prime}+f\left(  y\right)  y^{\prime2}+g(y)y^{\prime}+h(y)=0,
\label{eq2-2}%
\end{equation}
where $y^{\prime}=\frac{dy}{d\eta}=D_{\eta}y$, can be factorized in the form
\begin{equation}
\left[  D_{\eta}-\phi_{1}(y)y^{\prime}-\phi_{2}(y)\right]  \left[  D_{\eta
}-\phi_{3}(y)\right]  y=0, \label{eq2-3}%
\end{equation}
under the conditions%
\begin{align}
&  f\left(  y\right)  =-\phi_{1},\\
&  g(y)=\phi_{1}\phi_{3}y-\phi_{2}-\phi_{3}-\frac{d\phi_{3}}{dy}%
y,\label{eq2-4}\\
&  h(y)=\phi_{2}\phi_{3}y.
\end{align}

If we assume $\left[  D_{\eta}-\phi_{3}(y)\right]  y=\Omega(y)$, then the
factorized Eq. (\ref{eq2-3}) can be rewritten as
\begin{align}
y^{\prime}-\phi_{3}y  &  =\Omega,\label{eq3}\\
\Omega^{\prime}-\left(  \phi_{1}y^{\prime}+\phi_{2}\right)  \Omega &  =0.
\label{eq4}%
\end{align}

We can introduce the functions $\phi_{i}$ by comparing Eqs. (\ref{eq2}) and
(\ref{eq2-2}). Then, $\phi_{1}=\frac{A}{2y}$, $\phi_{2}=a_{1}^{-1}$ and
$\phi_{3}=2a_{1}(D+Ey^{\frac{1}{2}-q})$, where $a_{1}(\neq0)$ is an arbitrary
constant, are proposed.

Eq. (\ref{eq4}) can be easily solved for the chosen factorizing functions
obtaining as result $\Omega=\kappa_{1}e^{\eta/a_{1}}y^{A/2}$, where
$\kappa_{1}$ is an integration constant. Then, Eq. (\ref{eq3}) turns into the
equation
\begin{equation}
y^{\prime}-2a_{1}\left(  D+Ey^{\frac{1}{2}-s}\right)  y-\kappa_{1}%
e^{\eta/a_{1}}y^{A/2}=0, \label{PALOMA}%
\end{equation}
whose solution is also solution of Eq. (\ref{eq2}).

Furthermore, the following relationship is obtained from Eq. (\ref{eq2-4}),
\begin{equation}
Aa_{1}D-a_{1}^{-1}-2a_{1}D + a_{1}E(A-3+2s)y^{\frac{1}{2}-s}= 3 + Cy^{\frac
{1}{2}-s}. \label{eq2-5}%
\end{equation}
Eq. (\ref{eq2-5}) is a noteworthy result which provides the explicit form of
$a_{1}$ and the relationship among the parameters entering Eq. (\ref{eq2}).
Then, the viscous parameter $s$ as a function of parameter $\gamma$ is
obtained. By comparing both sides of Eq. (\ref{eq2-5}) and assuming
$r=1-\frac{1}{\gamma}$, leads to obtain:%
\begin{equation}
s\left(  \gamma\right)  _{\pm}=\frac{\pm\sqrt{2}+\gamma^{3/2}}{2\gamma^{3/2}}.
\label{lisa1}%
\end{equation}
Then, $s_{-}\in\lbrack0,.25]$ $\forall\gamma\in\lbrack1.2599,2]$, and
$s_{+}\in(.75,1.2071068]$ $\forall\gamma\in\lbrack1,2)$. Also, the explicit
form of $a_{1}$ is
\begin{equation}
a\left(  \gamma\right)  _{1\pm}=\pm\frac{2\gamma^{1/2}}{3(\sqrt{2}\mp
\gamma^{1/2})}. \label{lisa2}%
\end{equation}
Then, $a_{1-}\in\lbrack-1/3,-.29499]$ $\forall\gamma\in\lbrack1.2599,2]$, and
$a_{1+}\in\lbrack1.60947,\infty)$ $\forall\gamma\in\lbrack1,2)$.

We find the following significative values
\[%
\begin{array}
[c]{|c|c|c|c|c|}\hline
\gamma & s_{-} & a_{1-} & s_{+} & a_{1+}\\\hline\hline
1 &  &  & 1.2071 & 1.6095\\\hline
\frac{4}{3} & 4.0721\times10^{-2} & -0.299\,66 & 0.95928 & 2.9663\\\hline
2 & \frac{1}{4} & -\frac{1}{3} &  & \\\hline
\end{array}
\]
The main difference of these results from other approaches is expressed
through Eq. (\ref{lisa1}), which represents an advantage of the factorization
method as opposed to different approaches studied by other authors. This
equation provides the relationship between the parameters $s$ and $\gamma$ in
such a way that by fixing $s$ we get a particular value of $\gamma$.

The main dynamical variables of the FE are given in parametric form as
follows
\begin{align}
f\left(  \eta\right)   &  =f_{0}\exp\left(  \eta-\eta_{0}\right)
,\label{para1}\\
H\left(  \eta\right)   &  = y^{1/2}\left(  \eta\right)  ,\\
q(\eta)  &  = y^{1/2}\left(  \eta\right)  \frac{d}{d\eta}\left(  \frac
{1}{H\left(  \eta\right)  }\right)  -1,\\
\rho\left(  \eta\right)   &  = 3y\left(  \eta\right)  ,\\
p\left(  \eta\right)   &  = 3\left(  \gamma-1\right)  y\left(  \eta\right)
,\\
\Pi\left(  \eta\right)   &  = -\left(  3\gamma y\left(  \eta\right)
+\frac{dy}{d\eta}\right)  ,\\
l\left(  \eta\right)   &  = \frac{\left\vert \Pi\right\vert }{p},\\
\Sigma\left(  \eta\right)   &  = -3k_{B}\int\Pi(\eta)f^{3}(\eta)H(\eta
)T(\eta)^{-1}y(\eta)^{-1/2}d\eta. \label{para2}%
\end{align}

The authors have not been able to find the most general solution of
Eq. (\ref{PALOMA}). However, this equation can be studied for some
specific cases providing particular solutions of physical interest.
In Sections III and IV, the cosmological solutions as obtained for
the viscosity parameter $s\neq1/2$ and $s=1/2$ are studied.

\section{Solution with $s\neq1/2$.}

In this section, some particular cases of Eq. (\ref{PALOMA}) for $s\neq1/2$
are studied to obtain exact particular solutions of FE (\ref{fe1}%
)-(\ref{fe4}). By setting $\kappa_{1}=0$, Eq. (\ref{PALOMA}) simplifies as
\begin{equation}
y^{\prime}-2a_{1}\left(  D+Ey^{\frac{1}{2}-s}\right)  y=0, \label{PALOMA2}%
\end{equation}
whose solution is given by%
\begin{equation}
y(\eta)=\left(  \kappa_{2}e^{a_{1}D(2s-1)\eta}-\frac{E}{D}\right)
^{2/(2s-1)}, \label{sec3-eq2}%
\end{equation}
where $\kappa_{2}$ is an integration constant. Therefore, the parametric form
of the time function is obtained from Eq. (\ref{cv1}) as follows%
\begin{equation}
t\left(  \eta\right)  =\int y^{-1/2}(\eta)d\eta=\int\left(  \kappa_{2}%
e^{a_{1}D(2s-1)\eta}-\frac{E}{D}\right)  ^{1/(1-2s)}d\eta. \label{sec3-eq3}%
\end{equation}

\subsection{Case $s=0$.}

The first special case considered corresponds to $s=0$, which means that the
bulk viscosity coefficient $\xi=const$. The following particular solution is
obtained
\begin{equation}
y(\eta)=\left(  \kappa_{2}e^{-a_{1}D\eta}-\frac{E}{D}\right)  ^{-2}%
,\qquad\text{and\qquad}t\left(  \eta\right)  =-\frac{1}{Da_{1}}\left(
\kappa_{2}e^{-\eta Da_{1}}+\eta Ea_{1}\right)  . \label{eq3-13}%
\end{equation}

\begin{figure}[h!]
\begin{center}
\includegraphics[height=1.2228in,width=5.9352in]{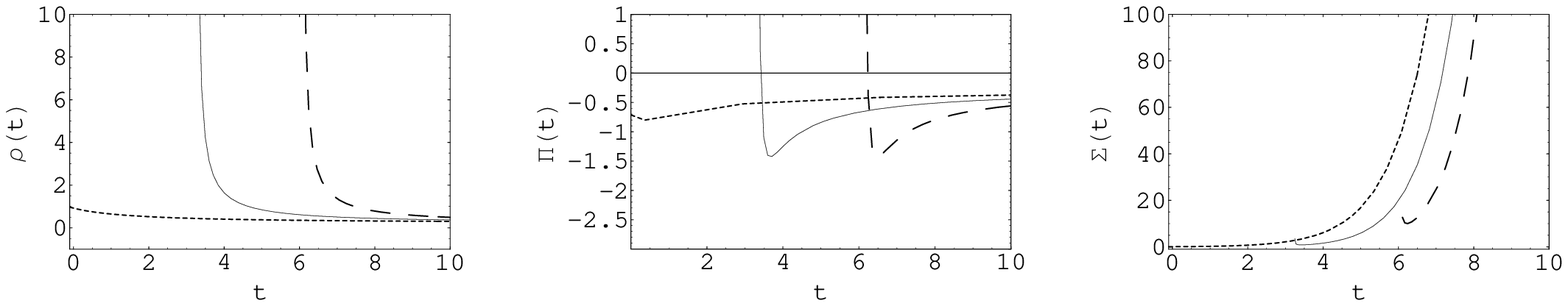}%
\caption{Solution with $s=0.$ Plots of energy density $\rho(t)$,
bulk viscosity $\Pi(t)$ and entropy $\Sigma(t)$. Dashed line for
$\kappa_{2}=-1.$ Solid line for $\kappa_{2}=-2.$
Long dashed line for $\kappa_{2}=-3.$}%
\label{sec3bpic1}%
\end{center}
\end{figure}

\begin{figure}[h!]
\begin{center}
\includegraphics[height=1.211in,width=4.7684in]{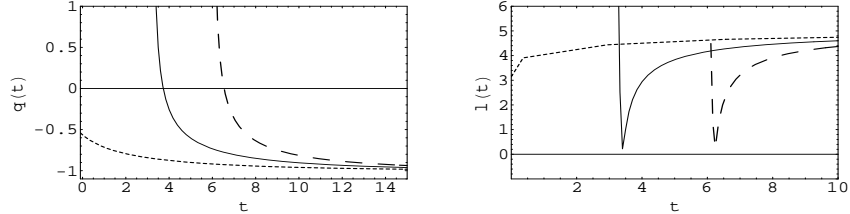}%
\caption{Solution with $s=0.$ Plots of the deceleration parameter
$q(t)$ and parameter $l(t)$. Plots of energy density $\rho(t)$, bulk
viscosity $\Pi(t)$ and entropy $\Sigma(t)$. Dashed line for
$\kappa_{2}=-1.$ Solid line for $\kappa_{2}=-2.$ Long dashed line
for $\kappa_{2}=-3.$}%
\label{sec3bpic2}%
\end{center}
\end{figure}

Eqs. (\ref{lisa1}) and (\ref{lisa2}) provide the corresponding
constant parameters $a_{1-}=-0.295$ and $\gamma=\sqrt[3]{2}$,
respectively. A particular equation of state is obtained once
again through Eq. (\ref{lisa1}). In Figs. \ref{sec3bpic1} and
\ref{sec3bpic2}, the behavior of the FE main quantities for
different values of constant $\kappa_{2}$ is plotted.

As we can see, the solution for $\kappa_{2}=-1$ is non-singular
since $\rho(0)=const$. For $\kappa_{2}=-2$ and $\kappa_{2}=-3$,
the energy density has a singular behavior when $t=0$, since it
runs to infinity when time tends to zero, i.e.,
$\rho(0)\rightarrow\infty$. The bulk viscosity, $\Pi$, is negative
for all values of $t$, i.e., $\Pi\left(  t\right)  <0$ $\forall
t\in\mathbb{R}^{+},$ which is a thermodynamically consistent
result as expected for $\kappa_{2}=-1$. For $\kappa_{2}=-2$ and
$\kappa_{2}=-3,$ the solution is valid only when $t>t_{0},$ i.e.,
$\Pi\left(  t\right)  <0$ $\forall t>t_{0}$, while $\Pi\left(
t\rightarrow0\right)  >0$. Then, for this interval of time,
$t\in\left(  0,t_{0}\right)  $, the solution has no physical
meaning. The entropy behaves like a strictly growing time
function; then, there are a large amount of comoving entropy
during the expansion of the universe. The deceleration parameter
runs from $q(0)=-0.5$ to $q(t)=-1$. Then, the solution is
accelerating, i.e., it is inflationary. The deceleration parameter
tends to $-1$ as $t\rightarrow\infty$ (accelerating solutions) but
shows a singular behavior when time runs to zero. The parameter
$l(t)$ shows that all the plotted solutions are far from
equilibrium since they are inflationary solutions, which is a
consistent result. To the best of our knowledge this solution is
new.

\subsection{Case $s=1/4$.}

The second case considered corresponds to $s=1/4$. In this case, Eqs.
(\ref{lisa1}) and (\ref{lisa2}) provide $a_{1}=-1/3$ and $\gamma=2$.
Therefore, Eq. (\ref{PALOMA}) simplifies as
\begin{equation}
y^{\prime}+2(3)^{3/4}y^{5/4}-\kappa_{1}e^{-3\eta}y^{3/4}=0. \label{berta}%
\end{equation}
If we perform the transformation $z=y^{1/4}$ in Eq. (\ref{berta}), then we get
the Riccati equation
\begin{equation}
z^{\prime}+\frac{3^{3/4}}{2}z^{2}-\frac{1}{4}\kappa_{1}e^{-3\eta}=0,
\label{eq3-15}%
\end{equation}
whose general solution is given in terms of Bessel $J_{n}$ and Neumman $N_{n}$
functions,
\begin{equation}
z(\eta)=-\xi(\eta)\frac{J_{1}(\xi(\eta))+\kappa_{2}N_{1}(\xi(\eta))}{J_{0}%
(\xi(\eta))+\kappa_{2}N_{0}(\xi(\eta))}, \label{eq3-16}%
\end{equation}
where $\xi(\eta)=\frac{\sqrt{2\kappa_{1}}}{2\cdot3^{3/8}}e^{-3\eta/2}$ and
$\kappa_{2}$ is an integration constant. Therefore, the following special
solution for Eq. (\ref{berta}) is obtained:
\begin{equation}
y(\eta)=\left(  \xi(\eta)\frac{J_{1}(\xi(\eta))+\kappa_{2}N_{1}(\xi(\eta
))}{J_{0}(\xi(\eta))+\kappa_{2}N_{0}(\xi(\eta))}\right)  ^{4},\quad
t(\eta)=\int^{\eta}\left(  \xi(\eta)\frac{J_{1}(\xi(\eta))+\kappa_{2}N_{1}%
(\xi(\eta))}{J_{0}(\xi(\eta))+\kappa_{2}N_{0}(\xi(\eta))}\right)  ^{-2}d\eta.
\label{eq3-17}%
\end{equation}

\begin{figure}[h!]
\begin{center}
\includegraphics[height=1.4981in,width=6.294in]{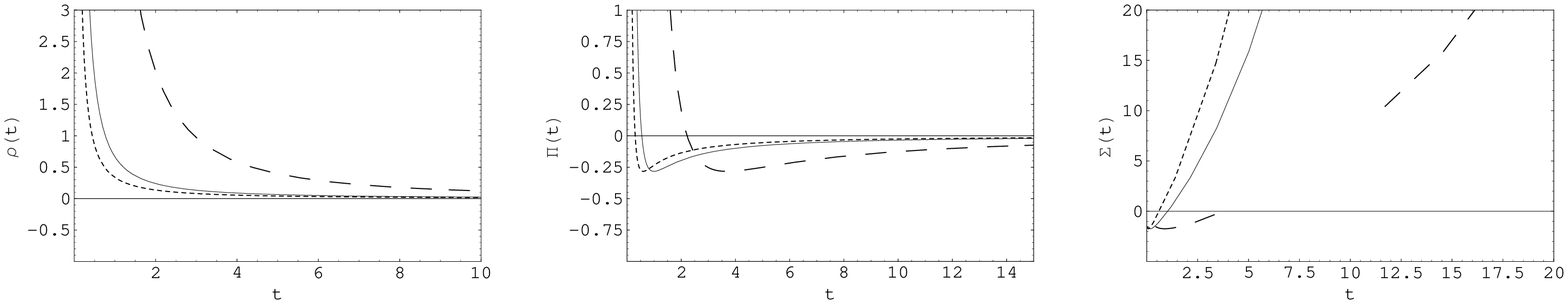}%
\caption{Solution with $s=1/4$ and $\gamma=2.$ Plots of energy
density $\rho(t)$, bulk viscosity $\Pi(t)$ and entropy $\Sigma(t)$.
Dashed line for $\kappa_{1}=4$, $\kappa_{2}=-10.$ Solid line for
$\kappa_{1}=19$, $\kappa_{2}=-3.$ Long
dashed line for $\kappa_{1}=3$, $\kappa_{2}=-0.7.$}%
\label{sec3c1pic1}%
\end{center}
\end{figure}

In order to study the behavior of the FE dynamical variables in
their parametric form, the calculation of Eq. (\ref{eq3-17}) has
been numerically addressed. The solution depends strongly on the
value of the numerical constants, in such a way that our solution is
physical only for $\kappa_{2}<0$ and for negative and relatively
small values ($<20$) of $\kappa_{1}$. Numerical analysis of the
solution plotted in Fig. \ref{sec3c1pic1} shows that the solution is
singular since the energy density tends to infinity when
$t\rightarrow0.$ The bulk viscosity is positive, $\Pi>0$, in the
region $\left( 0,t_{\ast}\right)  $ so the solution has physical
meaning only when $t>t_{\ast}$, for this era $\Pi$ becomes negative
as expected from the thermodynamical point of view and tending to
zero in the large time limit. In the same interval of time $\left(
0,t_{\ast }\right) $ the entropy production is negative,
$\Sigma(t)<0$ (unphysical situation), nevertheless when
$t>t_{\ast},$ a large amount of comoving entropy
is produced during the expansion of the universe.%

\begin{figure}[h]
\begin{center}
\includegraphics[height=1.5416in,width=5.1738in]{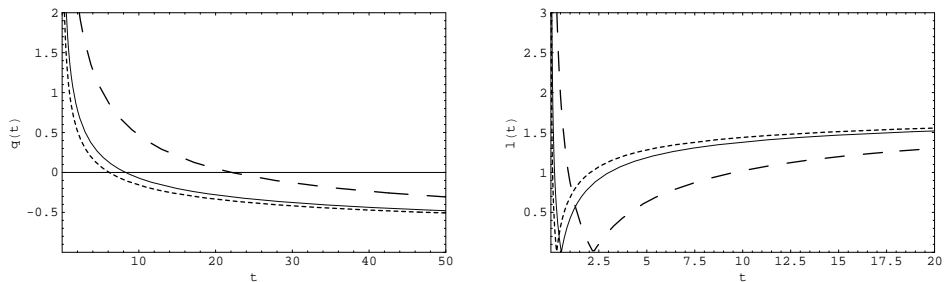}%
\caption{Solution with $s=1/4$ and $\gamma=2.$ Plots of the
deceleration parameter $q(t)$ and parameter $l(t)$. Dashed line for
$\kappa_{1}=4$, $\kappa_{2}=-10.$ Solid line for $\kappa_{1}=19$,
$\kappa_{2}=-3.$ Long
dashed line for $\kappa_{1}=3$, $\kappa_{2}=-0.7.$}%
\label{sec3c1pic2}%
\end{center}
\end{figure}

Regarding the dynamical behavior of solution (\ref{eq3-17}), in
Fig. \ref{sec3c1pic2} the behavior of parameters $q$ and $l$ has
been plotted. As we can see, the deceleration parameter shows that
the universe starts in a non-inflationary phase, but quickly
entering a inflationary one since $q<0.$ The plots of $l(t)$ are
consistent with this behavior, showing that the solution starts in
a thermodynamical equilibrium but in a finite time they are far
from equilibrium since they are inflationary solutions.

A similar solution has been obtained by Mak et al \cite{H6} but, as we have
shown, our solution is qualitatively different, with a very different physical meaning.

\subsubsection{A particular solution for the case $s=1/4$.}

\begin{figure}[h!]
\begin{center}
\includegraphics[height=1.2986in,width=6.1554in]{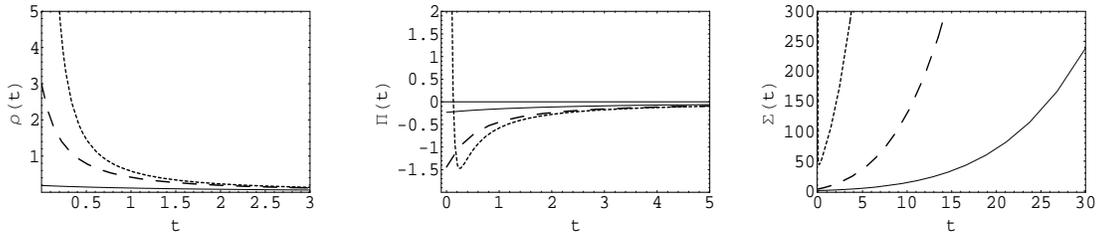}%
\caption{Particular solution for $s=1/4$. Plots of energy density
$\rho(t)$, bulk viscosity $\Pi(t)$ and entropy $\Sigma(t)$.
Dashed line for $\kappa_{2}=0.$ Long
dashed line for $\kappa_{2}=1.$ Solid line for $\kappa_{2}=2.$}%
\label{sec3cpic1}%
\end{center}
\end{figure}

\begin{figure}[h!]
\begin{center}
\includegraphics[height=1.1804in,width=4.2035in]{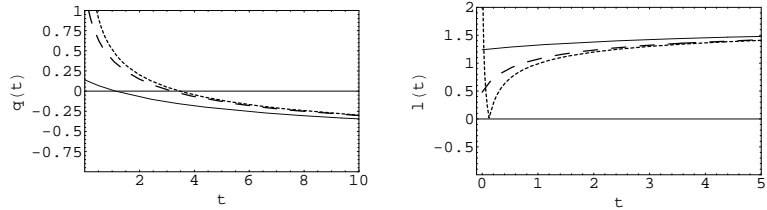}%
\caption{Particular solution for $s=1/4$. Plots of the deceleration
parameter $q(t)$ and parameter $l(t)$. Dashed line for
$\kappa_{2}=0.$ Long
dashed line for $\kappa_{2}=1.$ Solid line for $\kappa_{2}=2.$}%
\label{sec3cpic2}%
\end{center}
\end{figure}

If we set $\kappa_{1}=0$ in Eq. (\ref{berta}), then we get the very simple
ODE
\begin{equation}
y^{\prime}+2\left(  3\right)  ^{3/4}y^{5/4}=0,
\end{equation}
whose solution is given as
\begin{equation}
y(\eta)=\left(  \frac{\left(  3\right)  ^{3/4}}{2}\eta+\kappa_{2}\right)
^{-4},\qquad\text{and\qquad}t\left(  \eta\right)  =\frac{1}{4}\sqrt{3}\eta
^{3}+\frac{1}{2}3^{\frac{3}{4}}\eta^{2}\kappa_{2}+\eta\kappa_{2}^{2},
\end{equation}
where $\kappa_{2}$ is an integration constant. In Figs. \ref{sec3cpic1} and
\ref{sec3cpic2} the behavior of the FE main quantities has been plotted for
different values of the constant $\kappa_{2}.$%

The solution has been plotted for three different values of
constant $\kappa_{2}$. The energy density presents a singular
behavior only for $\kappa_{2}=0$, while the other two solutions
show a non-singular behavior when $t=0$. The solution for
$\kappa_{2}=2$ runs quickly to zero. The bulk viscosity is always
a negative time function for $\kappa_{2}=1$ and $\kappa_{2}=2$,
but the solution for $\kappa_{2}=0$ is valid only for $t>t_{0}$
since $\Pi(t\rightarrow0)>0$, which means that it lacks of
physical meaning in the interval of time $t\in\left(
0,t_{0}\right)  $. The entropy always behaves like a growing time
function but for the case $\kappa_{2}=0$ the universe starts with
a non-vanishing entropy, i.e., $\Sigma(0)=const.$, while for the
other two solutions $\Sigma(0)\rightarrow0.$ The plots in Fig.
\ref{sec3cpic1} show that a large amount of entropy is produced
during the expansion of the universe. Regarding the deceleration
parameter, the plotted solutions run to an acceleration region
since $q(t)\rightarrow-1/2$ in a finite time. For this reason, the
solution starts in an equilibrium regimen but quickly run to a
non-equilibrium state as shown by plots of $l(t)$. A particular
solution of this case has been studied by Harko et al \cite{H3}
obtaining different behavior of the FE main quantities.

\subsection{Case $s=1$.}

The second important case considered corresponds to $s=1$. According to Eqs.
(\ref{lisa1}) and (\ref{lisa2}), this solution is valid only for the equation
of state with $\gamma=\sqrt[3]{2}\thickapprox1.25992$.
Other authors have already studied similar cases for $s=1$, but with different
equation of state (see for instance \cite{H4} with $\gamma=2$) obtaining
different results. Then, according to Eqs. (\ref{sec3-eq2}) and
(\ref{sec3-eq3}), the following particular parametric solution is obtained:
\begin{equation}
y(\eta)=\left(  \kappa_{2}e^{a_{1}D\eta}-\frac{E}{D}\right)  ^{2}%
,\qquad\text{and}\qquad t\left(  \eta\right)  =\frac{1}{Ea_{1}}\left[
\ln\left(  -\frac{E}{Dk_{2}}+e^{\eta Da_{1}}\right)  -\eta Da_{1}\right]  .
\label{eq3-11}%
\end{equation}
Then, the FE main dynamical variables can be explicitly obtained through Eqs.
(\ref{para1})-(\ref{para2}).

\begin{figure}[h!]
\begin{center}
\includegraphics[height=1.4791in,width=6.3187in]{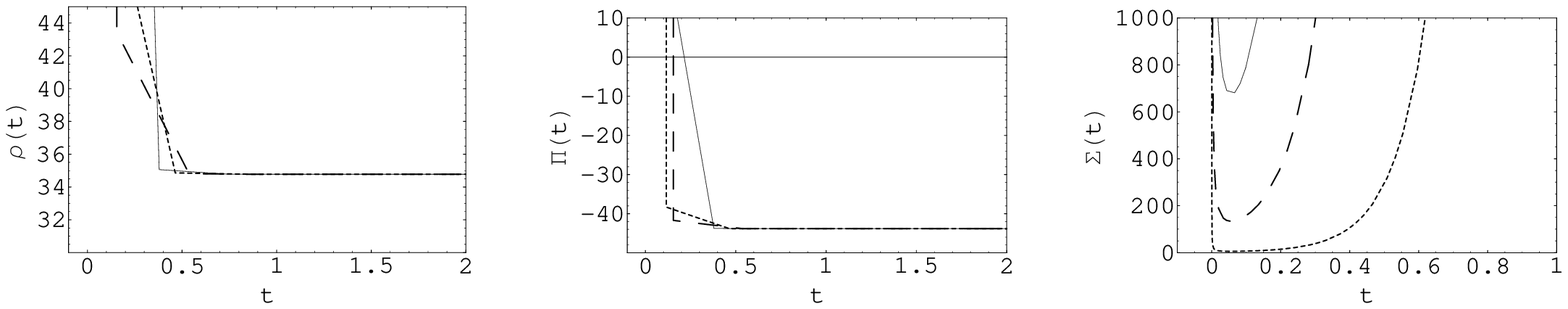}%
\caption{Solution with $s=1.$  Plots of energy density $\rho(t)$,
bulk viscosity $\Pi(t)$ and entropy $\Sigma(t)$. Dashed line for
$\kappa_{2}=0.1.$ Long dashed line for $\kappa_{2}=10.$ Solid line
for $\kappa_{2}=100.$}%
\label{sec3Apic1}%
\end{center}
\end{figure}

\begin{figure}[h!]
\begin{center}
\includegraphics[height=1.6063in,width=5.4389in]{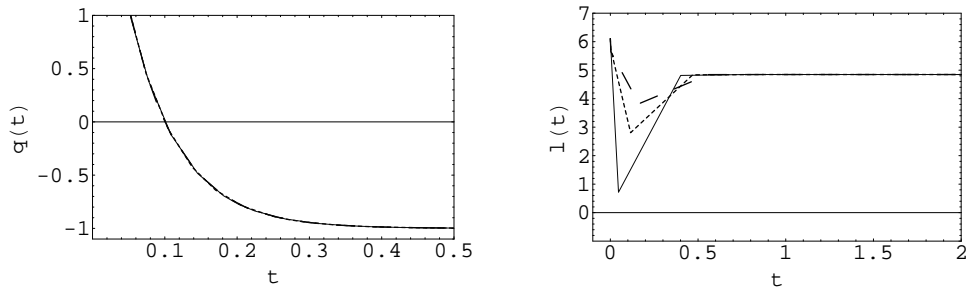}%
\caption{Solution with $s=1.$ Plots of the deceleration parameter
$q(t)$ and parameter $l(t)$. Dashed line for $\kappa_{2}=0.1.$ Long
dashed line for
$\kappa_{2}=10.$ Solid line for $\kappa_{2}=100.$}%
\label{sec3Apic2}%
\end{center}
\end{figure}

In Figs. \ref{sec3Apic1} and \ref{sec3Apic2}, the behavior of the main
quantities by giving different values to the constant $\kappa_{2}$ has been plotted.%

As we can see the solution is valid only for $t>t_{0}.$ The energy density is
a decreasing function, but the function behaves like a constant for a
$t>t_{c}$. The behavior of the bulk viscous parameter shows that the solution
is valid only for $t>t_{0}$ since the solution is positive when $t\rightarrow
0$, decreasing and going to a negative constant value during the cosmological
evolution, which is consistent from the thermodynamical point of view. In the
same way, the entropy behaves like a growing function only for $t>t_{0}$,
showing that a large amount of comoving entropy is produced. Nevertheless, the
deceleration parameter shows that the universe starts in a non-inflationary
phase, but quickly entering a inflationary one since $q\rightarrow-1$
$\forall\kappa_{2}.$ The plots of $l(t)$ show that plotted solutions are far
from equilibrium since they are inflationary solutions.

\section{Solution with $s=1/2$.}

We consider now the very special case $s=1/2$. This has been the most
important and studied case (see for example example \cite{C1},\cite{Ch97}%
,\cite{H1},\cite{H2}) within the framework of the bulk viscous cosmological
models since, as it has been pointed out for several authors, this solution is
stable from the dynamical systems point of view \cite{CoHoMa96} as well as
from the renormalization group approach \cite{TonyRG}.

In this case, Eq. (\ref{eq1}) reduces to:%
\begin{equation}
\ddot{H}-A_{1}\frac{\dot{H}^{2}}{H}+\left(  3+C_{1}\right)  H\dot{H}+\left(
D_{1}+E_{1}\right)  H^{3}=0, \label{nHarko1}%
\end{equation}
where $A_{1}=\left(  1+r\right)  =2-\frac{1}{\gamma}$, $C_{1}=\sqrt{3}$,
$D_{1}=\frac{9}{4}\left(  \gamma-2\right)  $, $E_{1}=\frac{3}{2}\sqrt{3}%
\gamma$, and $r=1-1/\gamma$. Since the coordinate transformation given by Eq.
(\ref{cv1}) leads to obtain several unphysical solutions for $s=1/2$, we
perform the more suitable change of variables given as follows (see also
\cite{C1}),
\begin{equation}
H=y^{1/2},\qquad d\eta=3\left(  1+\frac{1}{\sqrt{3}}\right)  Hdt.
\label{ncv_CH}%
\end{equation}
Then, Eq. (\ref{nHarko1}) turns into%
\begin{equation}
y^{\prime\prime}-\frac{A_{1}}{2y}y^{\prime2}+y^{\prime}+2\gamma by=0,
\label{nhelen1}%
\end{equation}
where $\gamma b=\frac{\sqrt{3}}{8}\left(  \gamma+6\right)  -\frac{3}{2}$. Eq.
(\ref{nhelen1}) can be solved by factorization providing new exact parametric
solutions for $s=1/2$.

Eq. (\ref{nhelen1}) admits the factorization
\begin{equation}
\left[  D-\frac{A}{2y}y^{\prime}-a_{1}^{-1}\right]  \left[  D-2a_{1}\gamma
b\right]  y=0,
\end{equation}
which can be rewritten in the form%
\begin{align}
y^{\prime}-2a_{1}\gamma by  &  =\Omega,\label{neq4}\\
\Omega^{\prime}-\left(  \frac{A}{2y}y^{\prime}-a_{1}^{-1}\right)  \Omega &
=0, \label{neq5}%
\end{align}
or equivalently,
\begin{equation}
y^{\prime}-2a_{1}\gamma by-\mathrm{k}_{1}e^{\eta/a_{1}}y^{A/2}=0,
\end{equation}
where $\mathrm{k}_{1}$ is an integration constant, with solution given as
\begin{equation}
y\left(  \eta\right)  =e^{2a_{1}\gamma b\eta}\left(  \frac{a_{1}\mathrm{k}%
_{1}e^{\left(  a_{1}^{-1}-a_{1}b\right)  \eta}}{2\gamma(1-a_{1}^{2}b)}%
+C_{1}\right)  ^{2\gamma}, \label{sol}%
\end{equation}
where $C_{1}$ is an integration constant, and the parameter $a_{1}$ is
restricted to values given by
\begin{equation}
a_{1\pm}=-\frac{4\sqrt{3}\gamma\pm\sqrt{\gamma^{2}\left(  72\sqrt
{3}-60\right)  -9\gamma^{3}+\gamma\left(  432\sqrt{3}-756\right)  }}{3\left(
\gamma-4\sqrt{3}+6\right)  }, \label{kyla}%
\end{equation}
i.e. , $a_{1+}\in\left[  -64.31,-8.38\right]  ,$ and $a_{1-}\in\left[
-0.23,-0.01\right]  $. In the following Subsections IV.A and IV.B several
possible cases of interest are studied.

\subsection{General solution.}

In this case it is possible to find a explicit parametric equation for $t$
(from Eq. (\ref{sol})) with $C_{1}\neq0.$ It is given as follows%
\begin{equation}
t\left(  \eta\right)  =\frac{\left(  \sqrt{3}-3\right)  a_{1}\left(
1+\frac{C_{1}\exp\left(  \frac{\eta}{2a_{1}\gamma}\left(  2\gamma
-a_{1}B\right)  \right)  }{a_{1}\mathrm{k}_{1}}\right)  ^{\gamma}\,}{6\gamma
y^{1/2}}\,_{2}F_{1}\left(  \frac{-2\gamma^{2}}{a_{1}B-2\gamma},\gamma
,1-\frac{2\gamma^{2}}{a_{1}B-2\gamma},-\frac{C_{1}\exp\left(  \frac{\eta
}{2a_{1}\gamma}\left(  2\gamma-a_{1}B\right)  \right)  }{a_{1}\mathrm{k}_{1}%
}\right)  .\label{sol_t}%
\end{equation}
To the best of our knowledge the solution given by Eqs. (\ref{sol}) and
(\ref{sol_t}) has not been previously reported.

\begin{figure}[h]
\begin{center}
\includegraphics[height=1.313in,width=6.5976in]{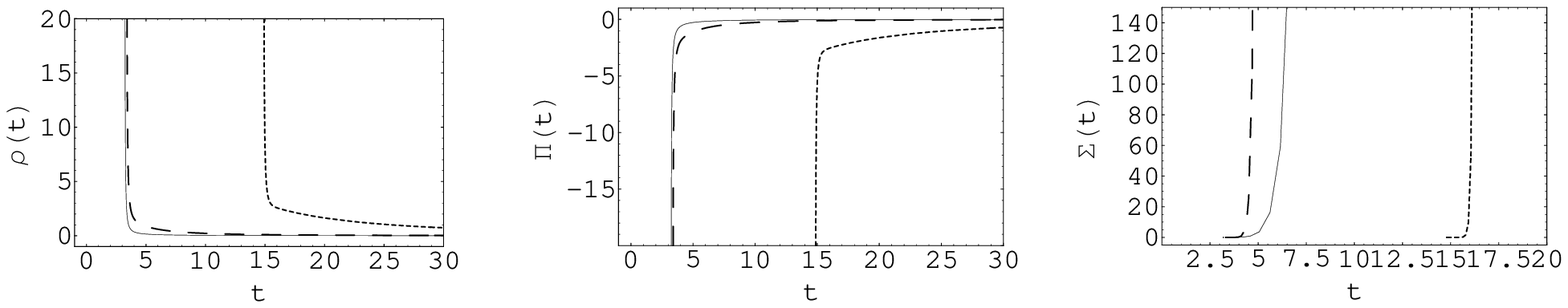}%
\caption{Solution for $s=1/2$. Plots of energy density $\rho(t)$,
bulk viscosity $\Pi(t)$ and entropy $\Sigma(t)$. Solid line for
$\gamma=2.$ Long dashed
line for $\gamma=4/3$. Dashed line for $\gamma=1.$}%
\label{nfpic1}%
\end{center}
\end{figure}

\begin{figure}[h]
\begin{center}
\includegraphics[height=1.295in,width=4.8992in]{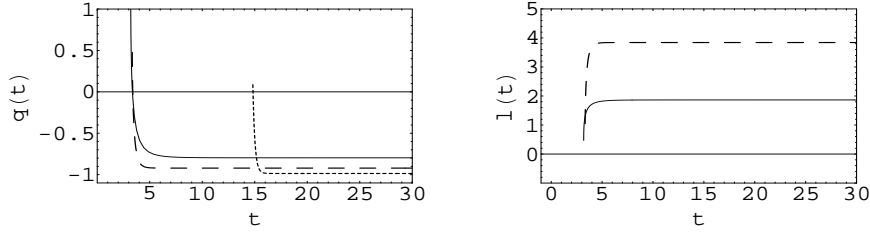}%
\caption{Solution for $s=1/2.$ Plots of the deceleration parameter
$q(t)$ and parameter $l(t)$. Solid line for $\gamma=2.$ Long
dashed line for $\gamma=4/3$. Dashed line for $\gamma=1.$}%
\label{nfpic2}%
\end{center}
\end{figure}

The FE main dynamical variables are given in parametric form as follows
\begin{align}
f\left(  \eta\right)   &  =f_{0}\exp\left(  \eta-\eta_{0}\right)  ,\\
H\left(  \eta\right)   &  =y^{1/2}\left(  \eta\right)  ,\\
q(\eta)  &  =-\frac{\left(  2+B\right)  C_{1}+2\left(  a_{1}+\gamma\right)
\mathrm{k}_{1}\exp\left(  \frac{\eta}{2a_{1}\gamma}\left(  2\gamma
-a_{1}B\right)  \right)  }{2\left(  C_{1}+a_{1}\mathrm{k}_{1}\exp\left(
\frac{\eta}{2a_{1}\gamma}\left(  2\gamma-a_{1}B\right)  \right)  \right)  },\\
\rho\left(  \eta\right)   &  =3y\left(  \eta\right)  ,\\
p\left(  \eta\right)   &  =3\left(  \gamma-1\right)  y\left(  \eta\right)  ,\\
\Pi\left(  \eta\right)   &  =\frac{C_{1}(B+3\gamma)+\left(  2+3a_{1}\right)
\gamma\mathrm{k}_{1}\exp\left(  \frac{\eta}{2a_{1}\gamma}\left(  2\gamma
-a_{1}B\right)  \right)  }{C_{1}+a_{1}\mathrm{k}_{1}\exp\left(  \frac{\eta
}{2a_{1}\gamma}\left(  2\gamma-a_{1}B\right)  \right)  }y\left(  \eta\right)
,\\
\Sigma\left(  \eta\right)   &  =\gamma e^{3\eta}\left(  3y\right)  ^{1/\gamma
},\\
l(\eta)  &  =\left\vert \frac{\Pi\left(  \eta\right)  }{p\left(  \eta\right)
}\right\vert .
\end{align}
In Figs. \ref{nfpic1} and \ref{nfpic2}, the behavior of the FE main quantities
has been plotted. The following constant values have been chosen: $a_{1+}$ as
given in Eq. (\ref{kyla}) while $B=\frac{\sqrt{3}a_{1}}{4}\left(
\gamma+6\right)  -3a_{1}$, $\mathrm{k}_{1}=2,\,$\ $C_{1}=-1$, and
$\gamma=1,4/3,2$ as usual. The solutions with $a_{1-}$ are unphysical.

The energy density shows a singular behavior as $t\rightarrow0$, but in a
finite time it behaves as a decreasing time function. This solution is valid
for all values of time and
$\gamma$. The bulk viscous pressure, $\Pi$, is a negative decreasing time
function during the cosmological evolution, $\Pi<0$ $\forall t\in
\mathbb{R}^{+}$, as it is expected from a thermodynamical point of view. The
viscous pressure also evolves from a singular era but it quickly tends to
zero, i.e., in the large limit the viscous pressure vanishes as the viscous
coefficient, which also becomes negligible small. The comoving entropy behaves
as a growing time function. There exists a fast growth of entropy for
$\gamma=4/3$, while for $\gamma=1$ the entropy grows slowly. The entropy
evolves from a non-singular state, i.e., $\Sigma(0)=0,$ but it quickly grows
in such a way that a large amount of entropy is produced during the
cosmological evolution. The picture of parameter $q(t)$ shows that all the
plotted solutions start in a non-inflationary phase, but they quickly run to
an inflationary era since this quantity runs to $-1$ for all the equations of
state. For this reason, the parameter $l(t)$ shows that the solutions are far
from equilibrium since they are inflationary solutions.

\subsection{Particular solution}

In the case, it is possible to find a particular solution for $t$ from Eq.
(\ref{sol}) with $C_{1}=0.$ For this case, the solution simplifies as follows%
\begin{equation}
y\left(  \eta\right)  =\exp\left(  B\eta\right)  \left(  \frac{a_{1}%
\mathrm{k}_{1}\exp\left(  \frac{\eta}{2a_{1}\gamma}\left(  2\gamma
-a_{1}B\right)  \right)  }{2\gamma-a_{1}B}\right)  ^{2\gamma},\text{\qquad
and\qquad}t\left(  \eta\right)  =\left(  \sqrt{3}-3\right)  \frac{a_{1}%
}{6\gamma}y^{-1/2}. \label{sc0}%
\end{equation}
Then, the FE main quantities are given in the following form:%
\begin{align}
f\left(  \eta\right)   &  =f_{0}\exp\left(  \eta-\eta_{0}\right)
,\label{sc1}\\
H\left(  \eta\right)   &  =y^{1/2}\left(  \eta\right)  ,\\
q(\eta)  &  =-\frac{\left(  a_{1}+\gamma\right)  }{a_{1}},\\
\rho\left(  \eta\right)   &  =3y\left(  \eta\right)  ,\\
p\left(  \eta\right)   &  =3\left(  \gamma-1\right)  y\left(  \eta\right)  ,\\
\Pi\left(  \eta\right)   &  =-\frac{\left(  2+3a_{1}\right)  \gamma}{a_{1}%
}y\left(  \eta\right)  ,\\
l\left(  \eta\right)   &  =\frac{1}{3}\left\vert \frac{\left(  2+3a_{1}%
\right)  \gamma}{a_{1}\left(  \gamma-1\right)  }\right\vert ,\\
\Sigma\left(  \eta\right)   &  =\gamma e^{3\eta}\left(  3y\left(  \eta\right)
\right)  ^{1/\gamma}, \label{sc8}%
\end{align}

It is possible to recover the known scaling solution studied by several
authors \cite{ZB},\cite{DZ} and \cite{Tony} from Eqs. (\ref{sc1})-(\ref{sc8}):%
\begin{align}
f  &  =f_{0}t^{H_{0}}\\
H\left(  t\right)   &  =H_{0}t^{-1},\\
q(t)  &  =H_{0}^{-1}-1,\\
\rho\left(  t\right)   &  =\rho_{0}t^{-2},\\
p\left(  t\right)   &  =3\left(  \gamma-1\right)  \rho_{0}t^{-2},\Pi\left(
t\right)  =-\Pi_{0}\rho\left(  t\right)  ,\\
l\left(  t\right)   &  =\frac{\Pi_{0}}{3\left(  \gamma-1\right)  },\\
\Sigma\left(  t\right)   &  \thickapprox\frac{\gamma\Sigma_{0}}{3\gamma
H_{0}-2}\left(  t^{\frac{1}{\gamma}\left(  3\gamma H_{0}-2\right)  }%
-t_{0}^{\frac{1}{\gamma}\left(  3\gamma H_{0}-2\right)  }\right)  ,
\end{align}
where $H_{0}=\frac{6\gamma}{\left(  \sqrt{3}-3\right)  a_{1}}$, $k_{B}^{-1}%
=1$, $\Sigma_{0}=-3\Pi_{0}H_{0}f_{0}^{3}\rho_{0}^{\frac{1}{\gamma}}\left(
\frac{1}{t_{0}}\right)  ^{3H_{0}}>0$, and $\Pi_{0}>0$.

\begin{figure}[h!]
\begin{center}
\includegraphics[height=1.1975in,width=6.2068in]{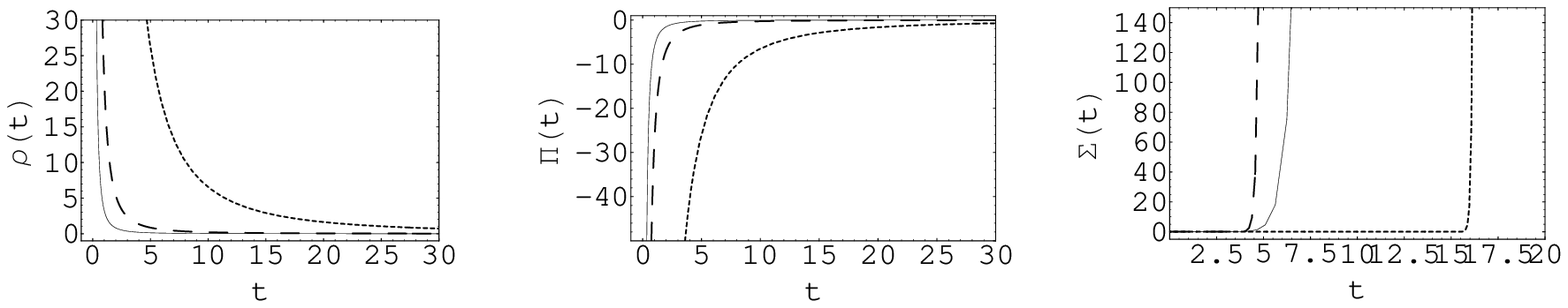}%
\caption{Solution for $s=1/2$ and $C_{1}=0$. Plots of energy density
$\rho(t)$, bulk viscosity $\Pi(t)$ and entropy $\Sigma(t)$. Solid
line for $\gamma=2.$ Long dashed line for $\gamma=4/3$. Dashed line
for $\gamma=1.$}%
\label{nfc2pic1}%
\end{center}
\end{figure}

\begin{figure}[h!]
\begin{center}
\includegraphics[height=1.1551in,width=4.1349in]{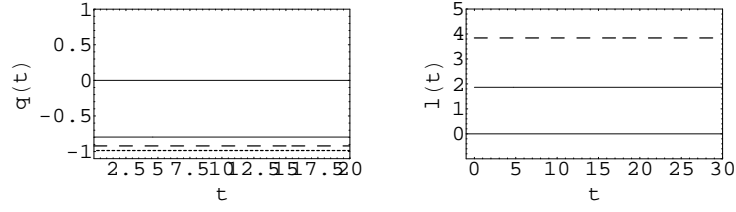}%
\caption{Solution for $s=1/2$ and $C_{1}=0$. Plots of the
deceleration parameter $q(t)$ and parameter $l(t)$. Solid line for
$\gamma=2.$ Long dashed line for $\gamma=4/3$. Dashed line for
$\gamma=1.$}%
\label{nfc2pic2}%
\end{center}
\end{figure}

In Figs. \ref{nfc2pic1} and \ref{nfc2pic2}, the FE main quantities have been
plotted using the same numerical values of Figs. \ref{nfpic1} and \ref{nfpic2}.%

All the plotted solutions have physical meaning $\forall$ $t$. These solutions
and the ones presented in the last solution (with $C_{1}\neq0)$ have a similar
behavior. We get the following numerical values for parameters $q(t)$ and
$l(t)$: $q_{1}=-0.98569$, $q_{4/3}=-0.92081$, $q_{2}=-0.796\,58$, while
$l_{4/3}=3.84$ and $l_{2}=1.86$.

\section{Solutions through the Lie group method for $s=1/2$}

In order to find new solutions and compare with the ones obtained through the
factorization method, we study the Hubble rate Eq. (\ref{nHarko1}) with
$s=1/2$ through Eq. (\ref{nhelen1}) by applying the Lie group method
\cite{Lie}. Eq. (\ref{nhelen1}) admits the following symmetries:
\begin{align}
\xi_{1}  &  = \left[  1,0\right]  ,\qquad\xi_{2}=\left[  0,y\right]
,\qquad\xi_{3}=\left[  1,y\right]  ,\nonumber\\
\xi_{4,5}  &  =\left[  0,y^{\frac{A}{2}}\exp\left(  \frac{\eta}{2}\left(  \pm
a-1\right)  \right)  \right]  ,\nonumber\\
\xi_{6,7}  &  =\left[  y^{\left(  1-\frac{A}{2}\right)  }\exp\left(
\frac{\eta}{2}\left(  1\mp a\right)  \right)  ,\frac{1\pm a}{A-2}\left(
y^{\left(  1-\frac{A}{2}\right)  }\exp\left(  \frac{\eta}{2}\left(  1\mp
a\right)  \right)  \right)  \right]  ,
\end{align}
where $a=\sqrt{1-8B+4AB}$, and $B=\frac{\sqrt{3}}{8}\left(  \gamma+6\right)
-\frac{3}{2}$. The non-zero constants $C_{ij}^{k}$ verifying the relationship
$\left[  \xi_{i},\xi_{j}\right]  =C_{ij}^{k}\xi_{k}$ are%
\begin{equation}
\left[  \xi_{1},\xi_{4}\right]  =C_{14}^{4}\xi_{4},\qquad\left[  \xi_{1}%
,\xi_{5}\right]  =C_{15}^{5}\xi_{5},\qquad\left[  \xi_{2},\xi_{4}\right]
=C_{24}^{4}\xi_{4},\qquad\left[  \xi_{2},\xi_{5}\right]  =C_{25}^{5}\xi_{5}.
\end{equation}
Then, we shall try to find a suitable change of variables with the symmetries
$\xi_{4}$ and $\xi_{5}$. These symmetries, $\xi_{4,5}=\left[  0,y^{A/2}%
e^{\eta/2\left(  \pm a-1\right)  }\right]  $, bring us to get the following cv
that will transform the original ODE into a quadrature. Following the the
standard procedure we get:
\begin{equation}
i=\eta,\,\qquad u(i)=\frac{1}{A-2}\left(  e^{\eta/2\left(  a\mp1\right)
}\left(  y^{1-A/2}\left(  \pm a-1\right)  +y^{A/2}y^{\prime}\left(
A-2\right)  \right)  \right)
\end{equation}
which lead us to obtain the following ODE and the corresponding solution:%
\begin{equation}
u^{\prime}=\mp au\qquad\Longrightarrow\qquad u=C_{1}e^{\mp ai}.
\end{equation}
Then, the solution to Eq. (\ref{nhelen1}) is given as follows%
\begin{equation}
y_{\mp}=\left(  \mp\frac{1}{2\gamma a}C_{1}e^{\frac{1}{2}\eta\left(  \mp
a-1\right)  }+C_{2}e^{\frac{1}{2}\eta\left(  \pm a-1\right)  }\right)
^{2\gamma} \label{sol_m}%
\end{equation}
where $a=\sqrt{1+4BA-8B}$, $A=2-\frac{1}{\gamma}$ and $B=\frac{\sqrt{3}}%
{8}\left(  \gamma+6\right)  -\frac{3}{2}$. In the following Subsections
V.A-V.D, the solutions provided in Eq. (\ref{sol_m}) are separately studied.

\subsection{Solution $y_{-}$ with $C_{2}\neq0$}

For the solution
\begin{equation}
y_{-}\left(  \eta\right)  =\left(  -\frac{1}{2\gamma a}C_{1}e^{\frac{1}{2}%
\eta\left(  -a-1\right)  }+C_{2}e^{\frac{1}{2}\eta\left(  a-1\right)
}\right)  ^{2\gamma}, \label{lg1}%
\end{equation}
with $C_{2}\neq0$, it is possible to find an explicit parametric equation for
$t$ through Eq. (\ref{ncv_CH}). It is given as%
\begin{equation}
t_{-}\left(  \eta\right)  =\frac{\left(  3-\sqrt{3}\right)  \left(
1-\frac{2a\gamma C_{2}e^{a\eta}}{C_{1}}\right)  ^{\gamma}\,}{3\left(
1+a\right)  \gamma y_{-}^{1/2}}\,_{2}F_{1}\left[  \gamma,\frac{\left(
1+a\right)  \gamma}{2a},\frac{\gamma+a\left(  2+\gamma\right)  }{2a}%
,\frac{2a\gamma C_{2}e^{a\eta}}{C_{1}}\right]  . \label{lg1a}%
\end{equation}
As we can see, a similar solution to the one obtained through the
factorization method has been found. However, as it is shown below, they
present several important differences.

\begin{figure}[h!]
\begin{center}
\includegraphics[height=1.5034in,width=6.1563in]{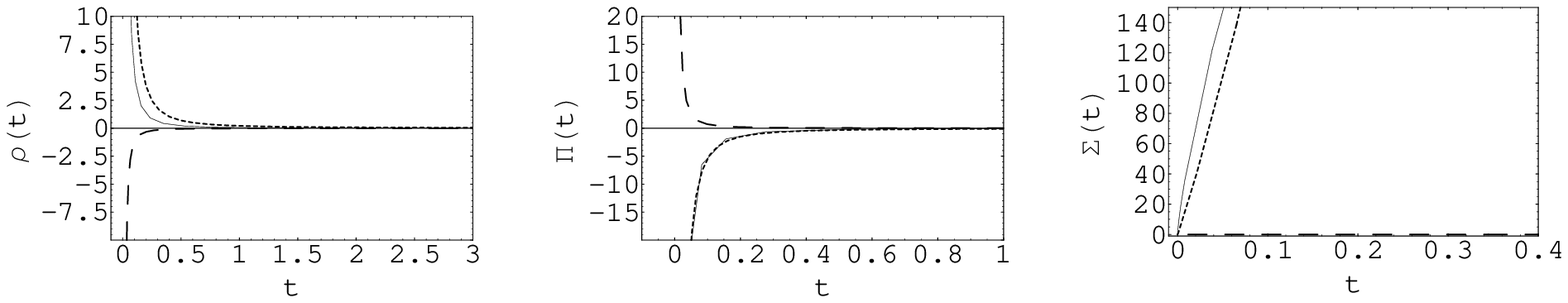}%
\caption{Solution for $y_{-}$ with $C_{2}\neq0$. Plots of energy
density $\rho(t)$, bulk viscosity $\Pi(t)$ and entropy
$\Sigma(t)$. Solid line for
$\gamma=2.$ Long dashed line for $\gamma=4/3$. Dashed line for $\gamma=1.$}%
\label{nfvgl1pic1}%
\end{center}
\end{figure}

\begin{figure}[h!]
\begin{center}
\includegraphics[height=1.4267in,width=5.1799in]{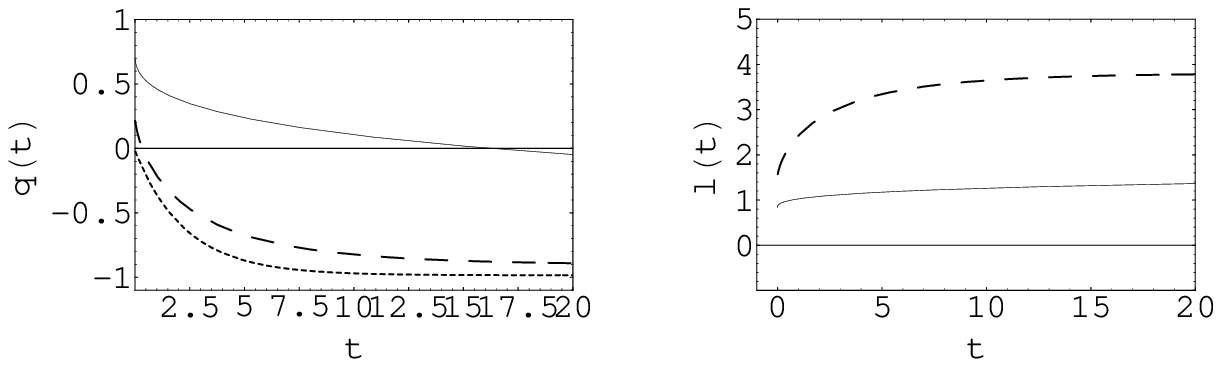}%
\caption{Solution for $y_{-}$ with $C_{2}\neq0$. Plots of the
deceleration parameter $q(t)$ and parameter $l(t)$. Solid line for
$\gamma=2.$ Long dashed line for $\gamma=4/3$. Dashed line for $\gamma=1.$}%
\label{nfvgl1pic2}%
\end{center}
\end{figure}

The FE main dynamical variables are given in parametric form as follows
\begin{align}
f\left(  \eta\right)   &  =f_{0}\exp\left(  \eta-\eta_{0}\right)  ,\\
H\left(  \eta\right)   &  =y_{-}^{1/2}\left(  \eta\right)  ,\label{eq3-5}\\
q(\eta)  &  =\frac{2a\gamma C_{2}e^{a\eta}\left(  2+\gamma\left(  a-1\right)
\right)  +C_{1}\left(  \gamma\left(  a+1\right)  -2\right)  }{2\left(
C_{1}-2a\gamma C_{2}e^{a\eta}\right)  },\\
\rho\left(  \eta\right)   &  =3y_{-}\left(  \eta\right)  ,\\
p\left(  \eta\right)   &  =3\left(  \gamma-1\right)  y_{-}\left(  \eta\right)
,\label{eq3-8}\\
\Pi\left(  \eta\right)   &  =\frac{\gamma\left(  2a\gamma\left(  a+2\right)
C_{2}e^{a\eta}+C_{1}\left(  a-2\right)  \right)  }{C_{1}-2a\gamma
C_{2}e^{a\eta}}y_{-}\left(  \eta\right)  ,\\
\Sigma\left(  \eta\right)   &  =\gamma e^{3\eta}\left(  3y_{-}\left(
\eta\right)  \right)  ^{1/\gamma},\\
l\left(  \eta\right)  )  &  =\frac{\left\vert \Pi\left(  \eta\right)
\right\vert }{p\left(  \eta\right)  } ,
\end{align}
In Figs. \ref{nfvgl1pic1} and \ref{nfvgl1pic2} the behavior of the FE main
quantities has been plotted.%

As it is shown in Fig. \ref{nfvgl1pic1}, the solution is not valid for
$\gamma=4/3$. For $\gamma=1$ (matter predominance) and $\gamma=2$ (ultra-stiff
matter), the energy density behaves as a decreasing time function during the
cosmological evolution. This solution is valid for all values of time, except
in the case $\gamma=4/3$, where $\rho_{4/3}<0$. The bulk viscosity is a
negative increasing time function, except in the case $\gamma=4/3$, where
$\Pi_{4/3}>0$. The energy-density, bulk viscosity and entropy have a very
similar behavior for the cases $\gamma=1$ and $\gamma=2$. The solution has a
singular origin since the energy density tends to infinity as $t\rightarrow0$.
The entropy is a growing time function which shows a large amount of comoving
entropy during the expansion of the universe. In the case $\gamma=4/3$, the
entropy starts growing at $t=60$, although we have ruled out this case. The
behavior of parameter $q(t)$ shows that the solution for $\gamma=2$ starts in
a non-inflationary phase, but after a period of time the solution enters an
inflationary era. Nevertheless, the solution for $\gamma=1$ is inflationary
for all values of $t$. The behavior of parameter $l(t)$ shows that the
solution for $\gamma=2$ is close to equilibrium, which is thermodynamically consistent.

\subsection{Solution $y_{-}$ with $C_{2}=0$}

\begin{figure}[h!]
\begin{center}
\includegraphics[height=1.5061in,width=7.0244in]{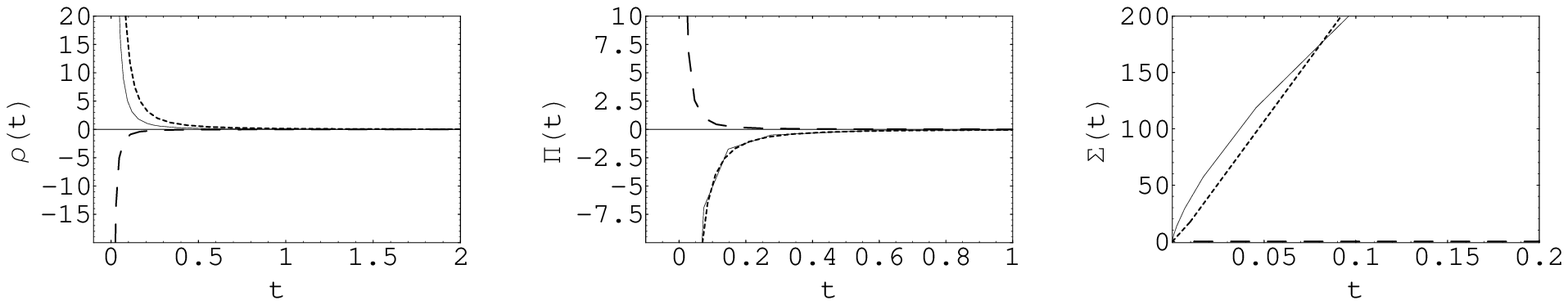}%
\caption{Solution $y_{-}$ with $C_{2}=0$. Plots of energy density
$\rho(t)$, bulk viscosity $\Pi(t)$ and entropy $\Sigma(t)$. Solid
line for
$\gamma=2.$ Long dashed line for $\gamma=4/3$. Dashed line for $\gamma=1.$}%
\label{nfvgl1c2pic1}%
\end{center}
\end{figure}

\begin{figure}[h!]
\begin{center}
\includegraphics[height=1.5666in,width=5.2638in]{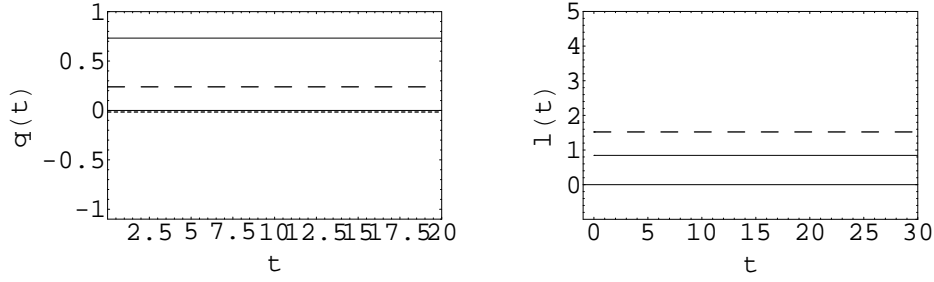}%
\caption{Solution $y_{-}$ with $C_{2}=0$. Plots of the
deceleration parameter $q(t)$ and parameter $l(t)$. Solid line for
$\gamma=2.$ Long dashed line for $\gamma=4/3$. Dashed line for $\gamma=1.$}%
\label{nfvgl1c2pic2}%
\end{center}
\end{figure}

For the case $y_{-}$ with $C_{2}=0$, the solution is given by (after simplifying)%
\begin{equation}
y_{-}\left(  \eta\right)  =\left(  -\frac{1}{2\gamma a}C_{1}e^{\frac{1}{2}%
\eta\left(  -a-1\right)  }\right)  ^{2\gamma},\qquad\text{and \qquad}%
t_{-}\left(  \eta\right)  =\int\left(  y_{-}\right)  ^{-1/2}d\eta
=\frac{\left(  3-\sqrt{3}\right)  }{3\left(  1+a\right)  \gamma}y_{-}^{-1/2}.
\label{lg4}%
\end{equation}
and the FE main dynamical variables are given in parametric form as follows
\begin{align}
f\left(  \eta\right)   &  =f_{0}\exp\left(  \eta-\eta_{0}\right)  ,\\
H\left(  \eta\right)   &  =y_{-}^{1/2}\left(  \eta\right)  ,\\
q(\eta)  &  =\frac{1}{2}\left(  \left(  1+a\right)  \gamma-2\right)  ,\\
\rho\left(  \eta\right)   &  =3y_{-}\left(  \eta\right)  ,\\
p\left(  \eta\right)   &  =3\left(  \gamma-1\right)  y_{-}\left(  \eta\right)
,\\
\Pi\left(  \eta\right)   &  =\left(  a-2\right)  \gamma y_{-}\left(
\eta\right)  ,\\
l\left(  \eta\right)   &  =\frac{1}{3}\left\vert \frac{\left(  a-2\right)
\gamma}{\left(  \gamma-1\right)  }\right\vert ,\\
\Sigma\left(  \eta\right)   &  =\gamma e^{3\eta}\left(  3y_{-}\left(
\eta\right)  \right)  ^{1/\gamma}.
\end{align}
The behavior of the FE main quantities has been plotted in Figs.
\ref{nfvgl1c2pic1} and \ref{nfvgl1c2pic2}. As it is observed, in this case, we
may recover the scaling solution.%

In this case, as in the last solution with $C_{2}\neq0$, the
solution for $\gamma=4/3$ is unphysical. All the main quantities
behave in the same way as the last solution with $C_{2}\neq0$
described above. Nevertheless, it is found that $q_{1}=-0.015$ for
$\gamma=1$ which represents an inflationary solution, and
$q_{2}=0.732$ for $\gamma=2$ which represents a non-inflationary
behavior, while $l_{2}=0.845<1$, i.e., the solution is within an
equilibrium regime. As it has been shown, most of the known exact
solutions of the gravitational FE with a viscous fluid do not
satisfy the condition $l<1$, i.e., the condition of thermodynamic
consistency, since they show an inflationary behavior. In the case
for $\gamma=2$, we have obtained a solution which is
thermodynamically consistent and it may describe the early
dynamics of a super-dense post-inflationary era when the
dissipative effects produced by the bulk viscosity may play an
important role.

\subsection{Solution $y_{+}$ with $C_{2}\neq0$}

\begin{figure}[h!]
\begin{center}
\includegraphics[height=1.4078in,width=6.7176in]{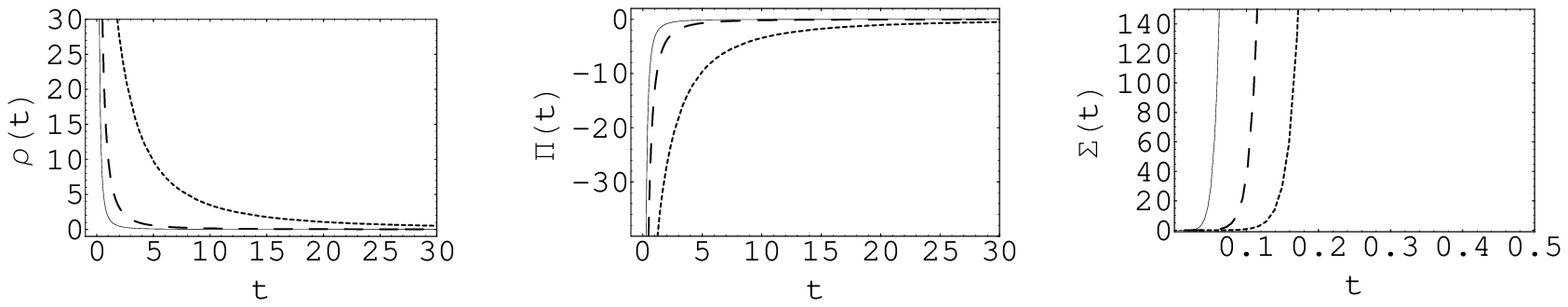}%
\caption{Solution $y_{+}$ with $C_{2}\neq0$. Plots of energy
density $\rho(t)$, bulk viscosity $\Pi(t)$ and entropy
$\Sigma(t)$. Solid line for $\gamma=2.$ Long dashed line for
$\gamma=4/3$. Dashed line for $\gamma=1.$}%
\label{nfvgl2pic1}%
\end{center}
\end{figure}

\begin{figure}[h!]
\begin{center}
\includegraphics[height=1.4655in,width=6.2086in]{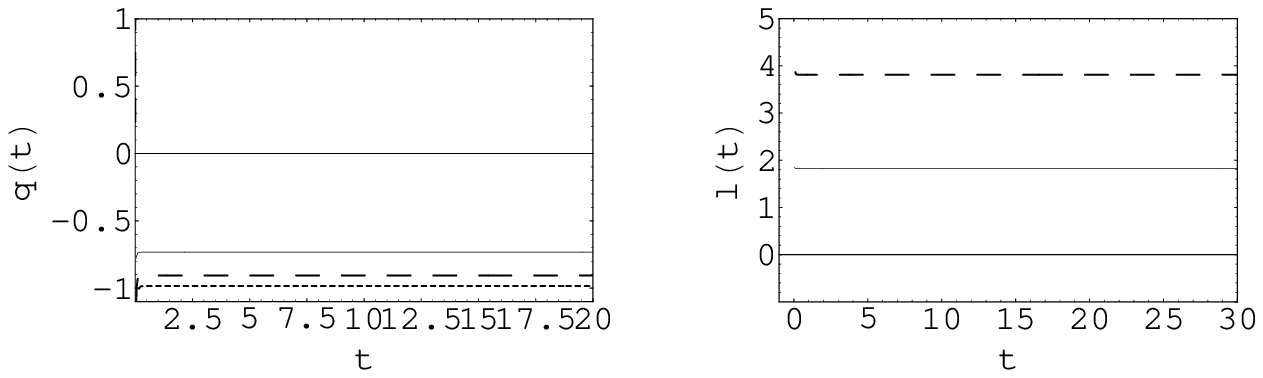}%
\caption{Solution $y_{+}$ with $C_{2}\neq0$. Plots of the
deceleration parameter $q(t)$ and parameter $l(t)$.  Solid line
for $\gamma=2.$ Long dashed line
for $\gamma=4/3$. Dashed line for $\gamma=1.$}%
\label{nfvgl2pic2}%
\end{center}
\end{figure}

For the solution given by
\begin{equation}
y_{+}\left(  \eta\right)  =\left(  \frac{1}{2\gamma a}C_{1}e^{\frac{1}{2}%
\eta\left(  a-1\right)  }+C_{2}e^{-\frac{1}{2}\eta\left(  a+1\right)
}\right)  ^{2\gamma},\label{lg2}%
\end{equation}
with $C_{2}\neq0$, we get the explicit parametric equation for the time
function
\begin{equation}
t_{+}\left(  \eta\right)  =\frac{\left(  3-\sqrt{3}\right)  }{3}\frac{\left(
1+\frac{C_{1}e^{a\eta}}{2a\gamma C_{2}}\right)  ^{\gamma}\,}{\left(
1+a\right)  \gamma y_{-}^{1/2}}\,_{2}F_{1}\left[  \gamma,\frac{\left(
1+a\right)  \gamma}{2a},\frac{\gamma+a\left(  2+\gamma\right)  }{2a}%
,-\frac{C_{1}e^{a\eta}}{2a\gamma C_{2}}\right]  .\label{lg2a}%
\end{equation}
The main dynamical variables of the FE are given in parametric form as
follows
\begin{align}
f_{+}\left(  \eta\right)   &  =f_{0}\exp\left(  \eta-\eta_{0}\right)  ,\\
H_{+}\left(  \eta\right)   &  =y_{+}^{1/2}\left(  \eta\right)  ,\\
q_{+}(\eta) &  =\frac{2a\gamma C_{2}\left(  \gamma\left(  a+1\right)
-2\right)  -C_{1}e^{a\eta}\left(  \gamma\left(  a-1\right)  +2\right)
}{2\left(  C_{1}-2a\gamma C_{2}e^{a\eta}\right)  },\\
\rho_{+}\left(  \eta\right)   &  =3y_{+}\left(  \eta\right)  ,\\
p_{+}\left(  \eta\right)   &  =3\left(  \gamma-1\right)  y_{+}\left(
\eta\right)  ,\\
\Pi_{+}\left(  \eta\right)   &  =\frac{\gamma\left(  2a\gamma\left(
a-2\right)  C_{2}-C_{1}\left(  a+2\right)  e^{a\eta}\right)  }{C_{1}e^{a\eta
}+2a\gamma C_{2}}y_{+},\\
l &  =\frac{1}{3}\left\vert \frac{\gamma\left(  2a\gamma\left(  a-2\right)
C_{2}-C_{1}\left(  a+2\right)  e^{a\eta}\right)  }{\left(  \gamma-1\right)
\left(  C_{1}e^{a\eta}+2a\gamma C_{2}\right)  }\right\vert ,\\
\Sigma_{+}\left(  \eta\right)   &  =\gamma e^{3\eta}\left(  3y_{+}\right)
^{1/\gamma}%
\end{align}
We have plotted the behavior of the FE main quantities in Figs.
\ref{nfvgl2pic1} and \ref{nfvgl2pic2}.%

This solution shows a behavior quite similar to the one obtained through the
factorization method. The energy density is a decreasing time function and it
is valid for all values of time. The bulk viscous pressure is a negative
increasing time function, while the entropy is a positive growing time
function. As in the case of the factorization method, the obtained solution is
valid for all the possible values of parameter $\gamma$. We find a fast growth
of entropy for $\gamma=2$, while it grows slowly during the evolution of the
universe for $\gamma=1$. The behavior of parameter $q(t)$ shows that all the
plotted solutions start in an inflationary phase, since this quantity is close
to $-1$ for every value of $\gamma$. The behavior of parameter $l(t)$ shows
that the solutions are far from equilibrium since these are inflationary solutions.

\subsection{Solution $y_{+}$ with $C_{2}=0$}

In the case of solution $y_{+}$ with $C_{2}=0$ we get
\begin{equation}
y_{+}\left(  \eta\right)  =\left(  \frac{1}{2\gamma a}C_{1}e^{\frac{1}{2}%
\eta\left(  a-1\right)  }\right)  ^{2\gamma},\qquad\text{and\qquad}%
t_{+}\left(  \eta\right)  =\frac{\left(  \sqrt{3}-3\right)  }{3\left(
a-1\right)  \gamma}y_{+}^{-1/2}. \label{lg3}%
\end{equation}
The FE main dynamical variables are given in parametric form as follows:
\begin{align}
f\left(  \eta\right)   &  =f_{0}\exp\left(  \eta-\eta_{0}\right)  ,\\
H\left(  \eta\right)   &  =y_{+}^{1/2}\left(  \eta\right)  ,\\
q(\eta)  &  =\frac{1}{2}\left(  \left(  1-a\right)  \gamma-2\right)  ,\\
\rho\left(  \eta\right)   &  =3y_{+}\left(  \eta\right)  ,\\
p\left(  \eta\right)   &  =3\left(  \gamma-1\right)  y_{+}\left(  \eta\right)
,\\
\Pi\left(  \eta\right)   &  =\left(  a+2\right)  \gamma y_{+},\\
l  &  =\frac{1}{3}\left\vert \frac{\left(  a+2\right)  \gamma}{\left(
\gamma-1\right)  }\right\vert ,\\
\Sigma\left(  \eta\right)   &  =\gamma e^{3\eta}\left(  3y_{+}\right)
^{1/\gamma}.
\end{align}
We may recover the scaling solution as above.%

\begin{figure}[h!]
\begin{center}
\includegraphics[height=1.4466in,width=5.8802in]{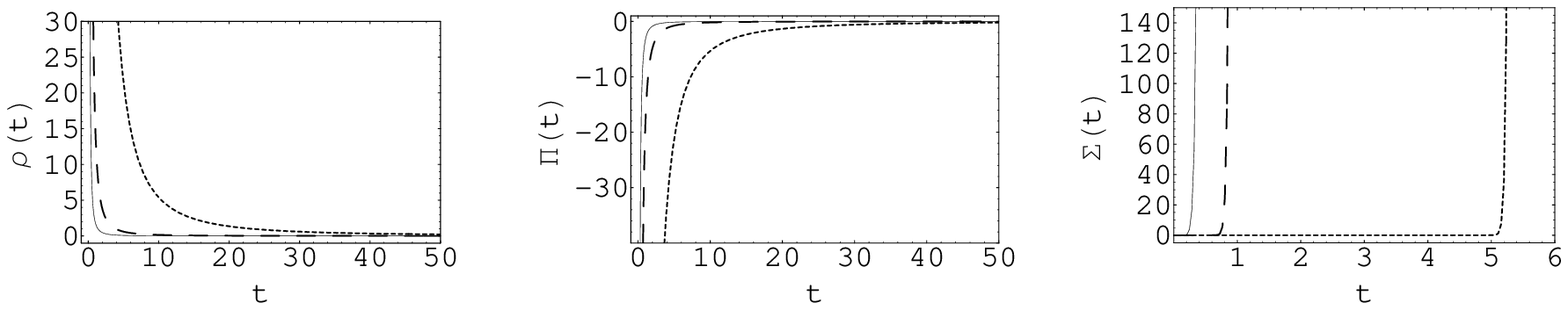}%
\caption{Solution $y_{+}$ with $C_{2}=0$. Plots of energy density
$\rho(t)$, bulk viscosity $\Pi(t)$ and entropy $\Sigma(t)$. Solid
line for $\gamma=2.$ Long dashed line for
$\gamma=4/3$. Dashed line for $\gamma=1.$}%
\label{nfvgl2c2pic1}%
\end{center}
\end{figure}

\begin{figure}[h!]
\begin{center}
\includegraphics[height=1.2697in,width=4.911in]{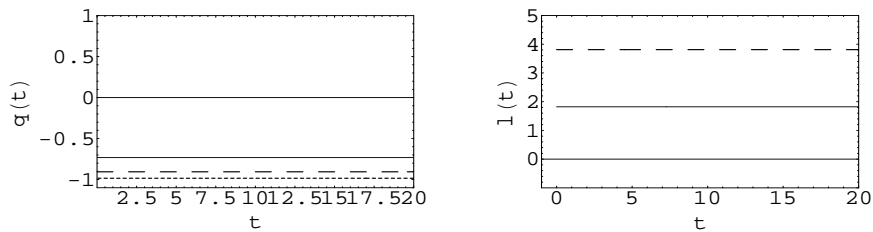}%
\caption{Solution $y_{+}$ with $C_{2}=0$. Plots of the
deceleration parameter $q(t)$ and parameter $l(t)$. Solid line for
$\gamma=2.$ Long dashed line for $\gamma=4/3$. Dashed line for $\gamma=1.$}%
\label{nfvgl2c2pic2}%
\end{center}
\end{figure}


In Figs. \ref{nfvgl2c2pic1} and \ref{nfvgl2c2pic2}, the behavior of the FE
main quantities has been plotted. As it can be seen, a very similar behavior
to the scaling solution obtained through the factorization method has been
obtained. Therefore, we get the same description and conclusions. It is worth
mentioning that the following values for the deceleration parameter $q(t)$ are
obtained: $q_{1}=-0.984206$, $q_{4/3}=-0.905604$, and $q_{2}=-0.732051$, while
for parameter $l(t)$ we obtain $l_{4/3}=3.81121$, and $l_{2}=1.82137$, i.e.,
the same values as the ones obtained for the scaling solution.

\section{Conclusions.}

In this work, we have studied a flat FRW cosmological model with a matter
model described as a full causal bulk viscous fluid. By assuming the state
equations given in Eq. (\ref{steq1}), the cosmological model simplifies to a
nonlinear second order ODE, the Hubble rate equation, for which a coordinate
transformation is performed in order to apply the factorization method. Due to
the coordinate transformation developed on the Hubble rate equation,
parametric exact solutions have been found. The standard procedure of
factorization provides the first order ODE (\ref{PALOMA}), and the restriction
condition given in Eq. (\ref{eq2-5}) which provides a relationship between the
viscous parameter $s$ and $\gamma$. Then, the analysis developed through
factorization allows to study the model for all the values of $s$ determined
by Eq. (\ref{lisa1}), instead of constructing a particular ODE for a single
given value of $s$ and arbitrary or specific values of $\gamma$, as it has
been previously studied by several authors.

We have studied several models for different values of $s$. Firstly, we have
studied and discussed the model for $s=0$, and $\gamma=\sqrt[3]{2}$. The
second model is studied for $s=1/4$, and $\gamma=2,$ finding two solutions.
The third case corresponds to $s=1$, and $\gamma=\sqrt[3]{2}\thickapprox
1.25992$. For the very special case $s=1/2$, the restriction equation
(\ref{eq2-5}) provides the explicit form of parameter $a_{1}$. However, the
obtained solutions have not restriction on the values of $\gamma$. For this
important case, we have been able to obtain a new solution which reduces, as
particular solution, to the known scaling solution. To the best of our
knowledge, the parametric solutions obtained for all these cases are new.

In order to obtain more new solutions, the case $s=1/2$ has been studied
through the Lie group method. The analysis carried out allows to obtain two
solutions. The solution (\ref{lg1})-(\ref{lg1a}) is new, and solution
(\ref{lg2})-(\ref{lg2a}) presents the same behavior as the one obtained
through the factorization method. Regarding the solution (\ref{lg1}%
)-(\ref{lg1a}), it is pointed out that it is not valid for all state equation
$\gamma$. It has been shown that for $\gamma=4/3$ the solution is unphysical,
while for $\gamma=2$ it is thermodynamically consistent and could be relevant
from the cosmological point of view.

\end{document}